\documentclass[aps,prl,twocolumn,print,groupedaddress,superscriptaddress,floatfix,10pt]{revtex4-1}
\usepackage{ulem}
\usepackage{graphicx}
\usepackage{soul,color}

\begin{document}

\title{Autonomous Scanning Probe Microscopy \textit{in-situ} Tip Conditioning \\through Machine Learning}

\author{Mohammad Rashidi}
\email[Correspondence to:]{ rashidi@ualberta.net}
\affiliation{Department of Physics, University of Alberta, Edmonton, Alberta, T6G 2J1, Canada.}
\affiliation{Quantum Silicon Inc., Edmonton, AB, Canada, T6G 2M9}

\author{Robert A. Wolkow}
\affiliation{Department of Physics, University of Alberta, Edmonton, Alberta, T6G 2J1, Canada.}
\affiliation{Quantum Silicon Inc., Edmonton, AB, Canada, T6G 2M9}
\affiliation{Nanotechnology Initiative, Edmonton, AB, Canada, T6G 2M9}

\begin{abstract}
Atomic scale characterization and manipulation with scanning probe microscopy rely upon the use of an atomically sharp probe. Here we present automated methods based on machine learning to automatically detect and recondition the quality of the probe of a scanning tunneling microscope. As a model system, we employ these techniques on the technologically relevant hydrogen-terminated silicon surface, training the network to recognize abnormalities in the appearance of surface dangling bonds. Of the machine learning methods tested, a convolutional neural network yielded the greatest accuracy, achieving a positive identification of degraded tips in 97\% of the test cases. By using multiple points of comparison and majority voting, the accuracy of the method is improved beyond 99\%. The methods described here can easily be generalized to other material systems and nanoscale imaging techniques.
\end{abstract}

\maketitle

The ability to directly visualize and manipulate individual atoms using scanning probe microscopy (SPM)~\cite{Schweizer1990,Crommie1993,Stroscio2004,Sugimoto2008,Kalff2016,Slot2016,Drost2017,Huff2017,Pavlicek2017a,Folsch2014,Kawai2014} has inspired scientists to develop atomic scale technology for over two decades. Among other things, these technologies can be used to create smaller, more efficient, faster and cheaper devices~\cite{Huff2017a,Khajetoorians2011}. To commercialize these technologies SPM fabrication must become fast, precise and automonous. To this end, several studies have demonstrated methods that make it feasible to build parallelized atomically precise robots that manipulate and analyze atoms automatically~\cite{Woolley2011,Castellano-Hernandez2012,Stirling2013,Moller2017,Ziatdinov2017,Ziatdinov2018}.
  
SPM techniques, and atomic manipulation in particular, rely on atomically sharp metal tips. While standard tip preparation methods can reliably produce them \textit{ex-situ} (for example, single atom tips can be prepared with field ion microscopy~\cite{Rezeq2006a}) during their use for imaging and atomic manipulation the quality of their apex is routinely compromised due to interactions with the surface. The signature of this is the loss of atomic resolution, or the appearance of secondary imaging features, indicating that the apex of the tip no longer has a single predominant atom. Such tips are  generally called ``double tips" for this reason. Because a single atom tip is required for SPM atomic fabrication and experiments, \textit{in-situ} tip treatments are necessary to return the tip to its ideal (sharp) condition. This is usually the most time-consuming process for SPM operators. Common methods include applying short voltage pulses between the tip and sample or controllably indenting the tip into the sample. These processes typically must be repeated many times before the tip's quality is restored.

Here, we use machine learning to automate \textit{in-situ} tip conditioning. Our process is based on a convolutional neural network (CNN) that is trained to analyze the quality of the tip. In our case, we work on the hydrogen terminated Si(100) substrate, a promising platform to develop atomic circuitry~\cite{Huff2017a,Livadaru2010,Rashidi2016b,Schofield2013,Fuechsle2012}. The network is trained to recognize and assess the image quality of isolated surface dangling bonds (~97\% accuracy). By using majority voting on a small set of dangling bond images the operational accuracy of the tip quality classification is improved beyond 99\%. Upon detection of degraded probe quality, the routine performs \textit{in-situ} tip conditioning on a preselected spot on the surface. This procedure is repeated until the network registers a sharp probe.

All experiments were performed on  an Omicron LT STM operating at 4.5~K and under ultrahigh vacuum. The tips were electrochemically etched from polycrystalline tungsten wire.  Tips were heated via electron bombardment in ultrahigh vacuum to remove the surface oxide, and sharpened to single atom by field ion microscopy~\cite{Rezeq2006a}. \textit{In-situ} tip processing was performed by controlled tip indentation with the surface. Samples are highly arsenic-doped (1.5$\times$10$^{19}$ atom/cm$^{3}$) Si(100). Samples were degassed at 600$^\circ$C for ~12 hours followed by flash annealing at 1250$^\circ$C. For hydrogen passivation, they are exposed to atomic hydrogen gas at 330$^\circ$C. A Nanonis SPM controller was used for imaging and data acquisition. The tip conditioning automation routine was programmed in Python and Labview using the Nanonis programming interface.  A k-nearest neighbor, random forest, support vector machine and fully connected neural network were implemented using the Scikit-Learn(0.19.1), Python machine learning library. The CNN was implemented using Keras(2.1.3) with TensorFlow backend.

\begin{figure*}[t]
	\includegraphics[width=175mm]{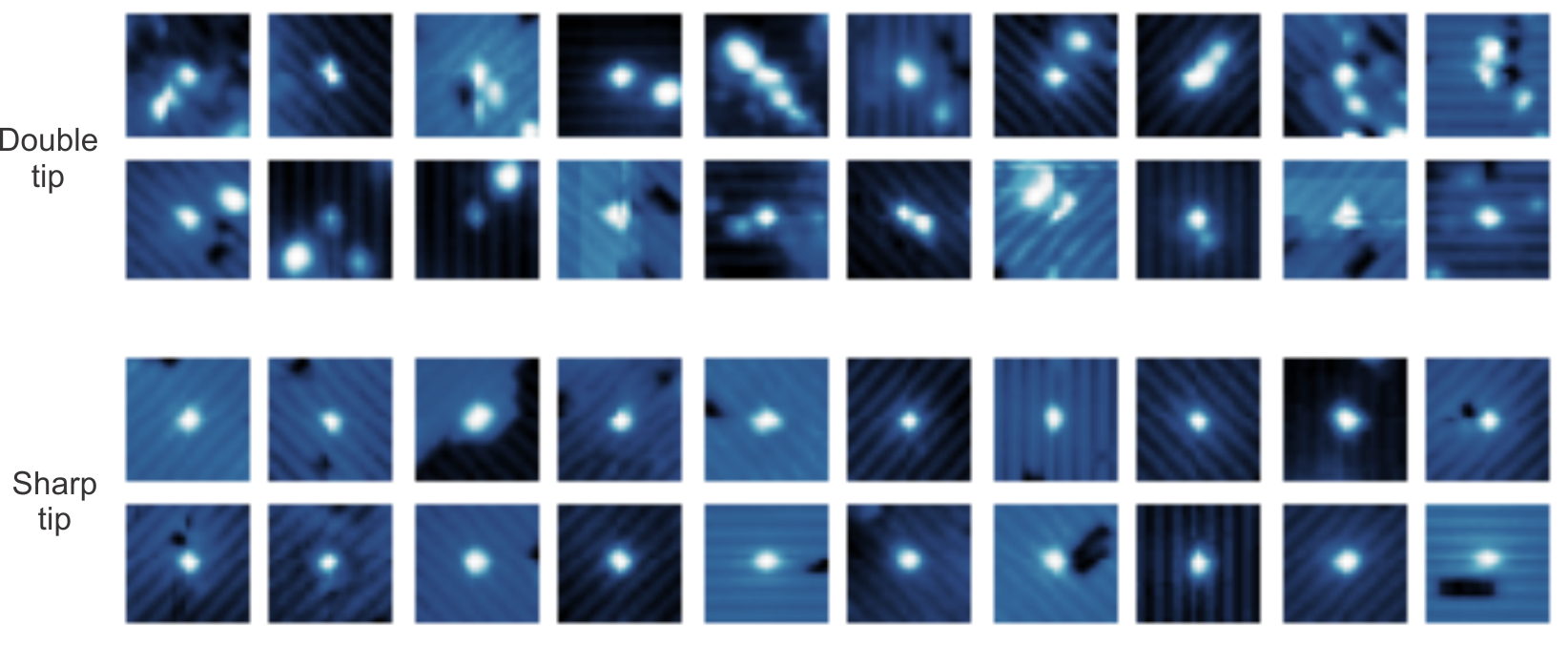}
	\caption{\label{Fig1} \textbf{Randomly selected labeled data used for training.} 5.6$\times$5.6~nm$^2$ dangling bond images are extracted from the STM images recorded at -1.8 V and 50 pA. }
\end{figure*}
For training, we used approximately 3500 STM images of isolated dangling bonds recorded at a sample bias of $-1.8$~V, where they typically appear as bright protrusions. These images were selected from five years of archived data from two of our microscopes. To enable direct comparison, each 5.6$\times$5.6~nm$^2$ image was resized to 28$\times$28 pixels. Each of the images was labeled manually (Fig.~1).  We augmented each image by rotating it by 90$^\circ$ four times and mirroring each rotated image. This expanded the training dataset by 8 times and resulted in a significant performance increase to each of the models we tested.

We tested several machine learning models on our dataset and selected a CNN model to implement in our automation routine because of its high precision score.
Table~I summarizes the outcome of each model. 
\begin{table}[h!]
\begin{tabular}{c||c|c|c|c|c}
	Model&KNN & RFC & SVM & FCNN & CNN \\
	\hline
	Precision  & 0.84 & 0.89 & 0.88 &0.78 &0.97   \\

\end{tabular}
\caption{Precision score of different machine learning models that have been tested for our dataset. KNN, RFC, SVM, FCNN and CNN denote K nearest neighbor, random forest classifier, support vector machine, fully connected neural network and convolutional neural network, respectively. KNN classifier with 5 neighbors resulted the best accuracy score for  our dataset. 5000 trees were used for RFC. The (Gaussian) radial basis function kernel with C and $\gamma$ parameters of 500 and 0.5 was used for SVM. The FCC had 18 hidden layers with ReLU activation function and Adam optimizer~\cite{Kingma2015} (learning rate of 10$^{-3}$).}
\label{table:1}
\end{table}

\begin{figure*}[t]
	\includegraphics[width=165mm]{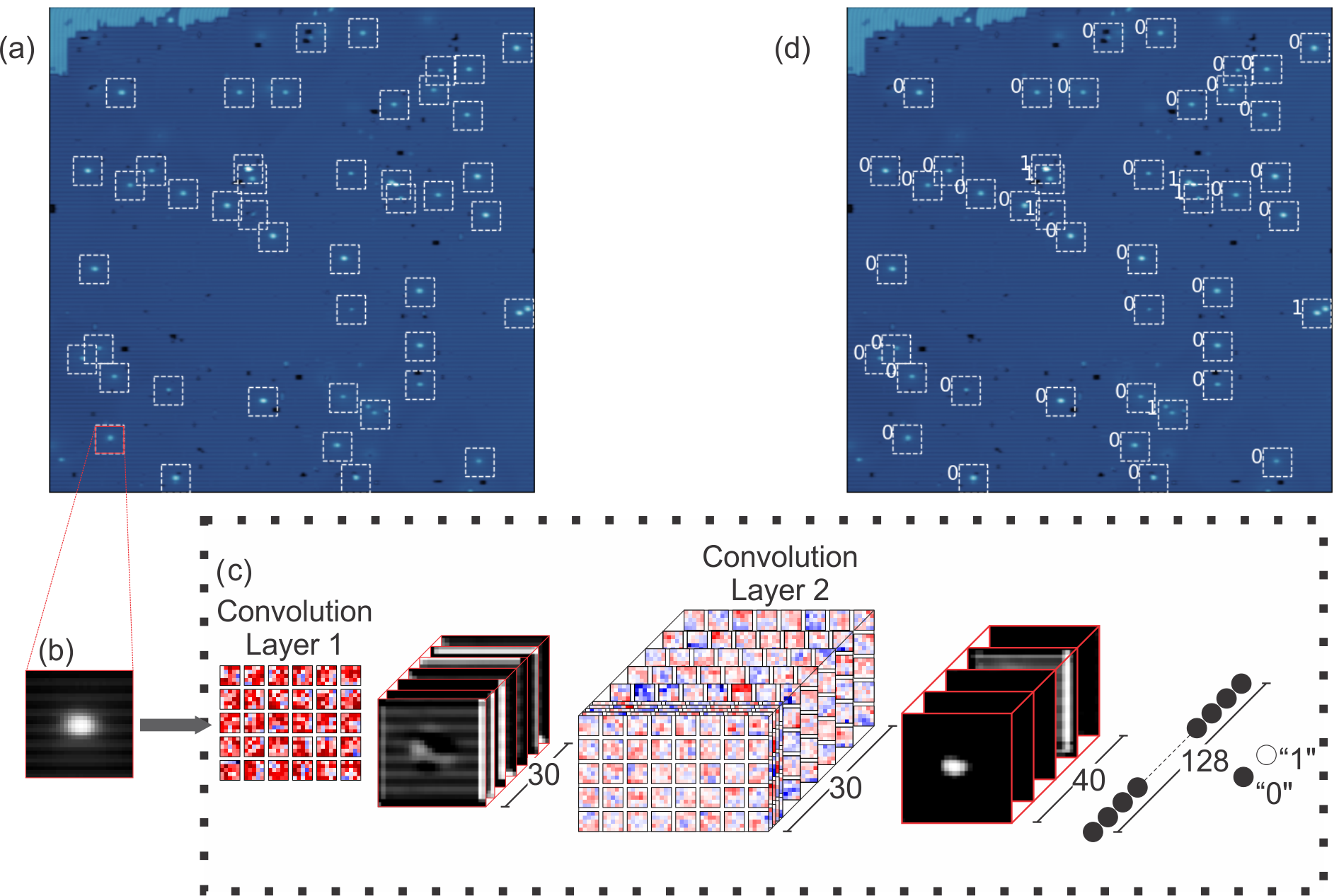}
	\caption{\label{Fig2} \textbf{Tip Quality Analysis with Convolutional Neural Network.} (a) STM image (100$\times$100 nm$^2$) of hydrogen-terminated Si(100) recorded at -1.80 V and 50 pA. Bright features are surface dangling bonds. The dangling bonds are automatically extracted from the image (black squares) and sequentially fed into the CNN. (b) Close up of the dangling bond indicated by the red square in (a). (c) A depiction of the CNN used in this study. It consists of two convolution layers followed by a pooling layer, a densely connected layer and an output layer. The result of the output layer is ``0" for sharp tips or ``1" for double tips. As an example, the output of each CNN layer is shown for the dangling bond image in (b). (d) The outputs of the CNN for each automatically extracted dangling bond image in (a). }
\end{figure*}

Figure~2 displays the workflow of the tip quality analysis using a CNN.  Our routine automatically identifies and isolates subsections of the STM image containing dangling bonds, and feeds them sequentially to the CNN to analyse the tip quality. The black squares in Fig. 2a indicate the dangling bond images that were used for analysis in this example. As an example, the output of the CNN for a dangling bond image in Fig. 2b is shown in Fig. 2c.  The CNN consists of two back to back 30 and 40 kernels (5$\times$5, stride 1) convolutional layers with  ReLU activation function. 
These layers were followed by a  max-pooling layer (2$\times$2, stride 2) flattened and fully connected to a 128 node layer with ReLU activation function.  A 2-node fully connected layer with Softmax activation function was used for classification at the end. The Adam optimizer~\cite{Kingma2015} with learning rate of 10$^{-4}$ and the categorical cross-entropy as loss function was used.
Figure 2d displays the output of the CNN for all the dangling bonds in the STM image. The program performs majority voting at the end to determine the outcome.

\begin{figure}
	\includegraphics[width=75mm]{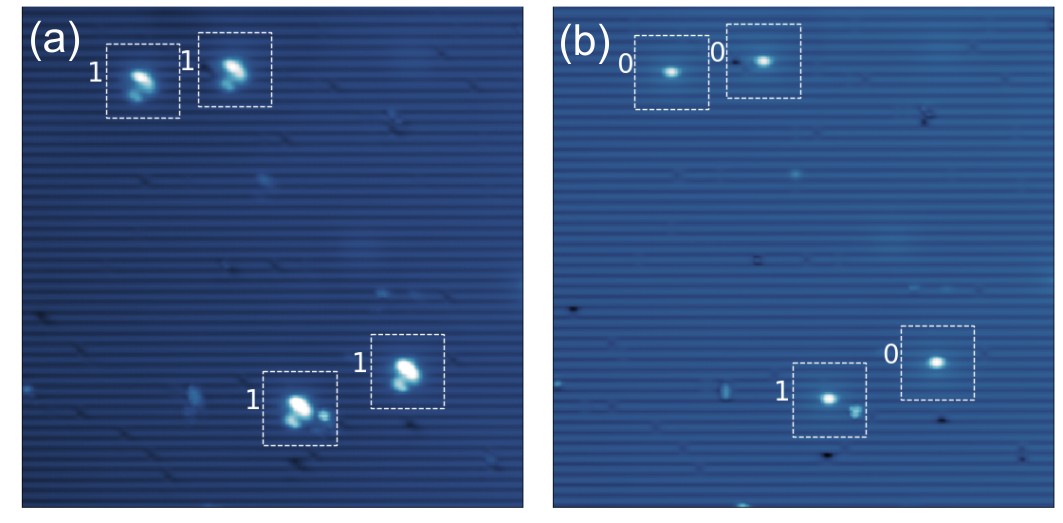}
	\caption{\label{Fig3} \textbf{An Example of Automatic Tip Sharpening.}  (a) Initial image to judge the quality of the tip. User sets an image frame and a spot for in-situ tip conditioning before starting the program.  The majority vote outputs of the CNN is ``1" (double tip) for this image and 4 other subsequent images (not shown here), indicating that the tip conditioning was not successful. (b) After a successful tip conditioning the majority vote output of the CNN becomes ``0" (sharp tip) and the program stops its operation. The size of the images were $40\times 40$~nm$^2$ and the tunneling conditions were $-1.8$~V and 50~pA.}
\end{figure}
\begin{figure*}
	\includegraphics[width=175mm]{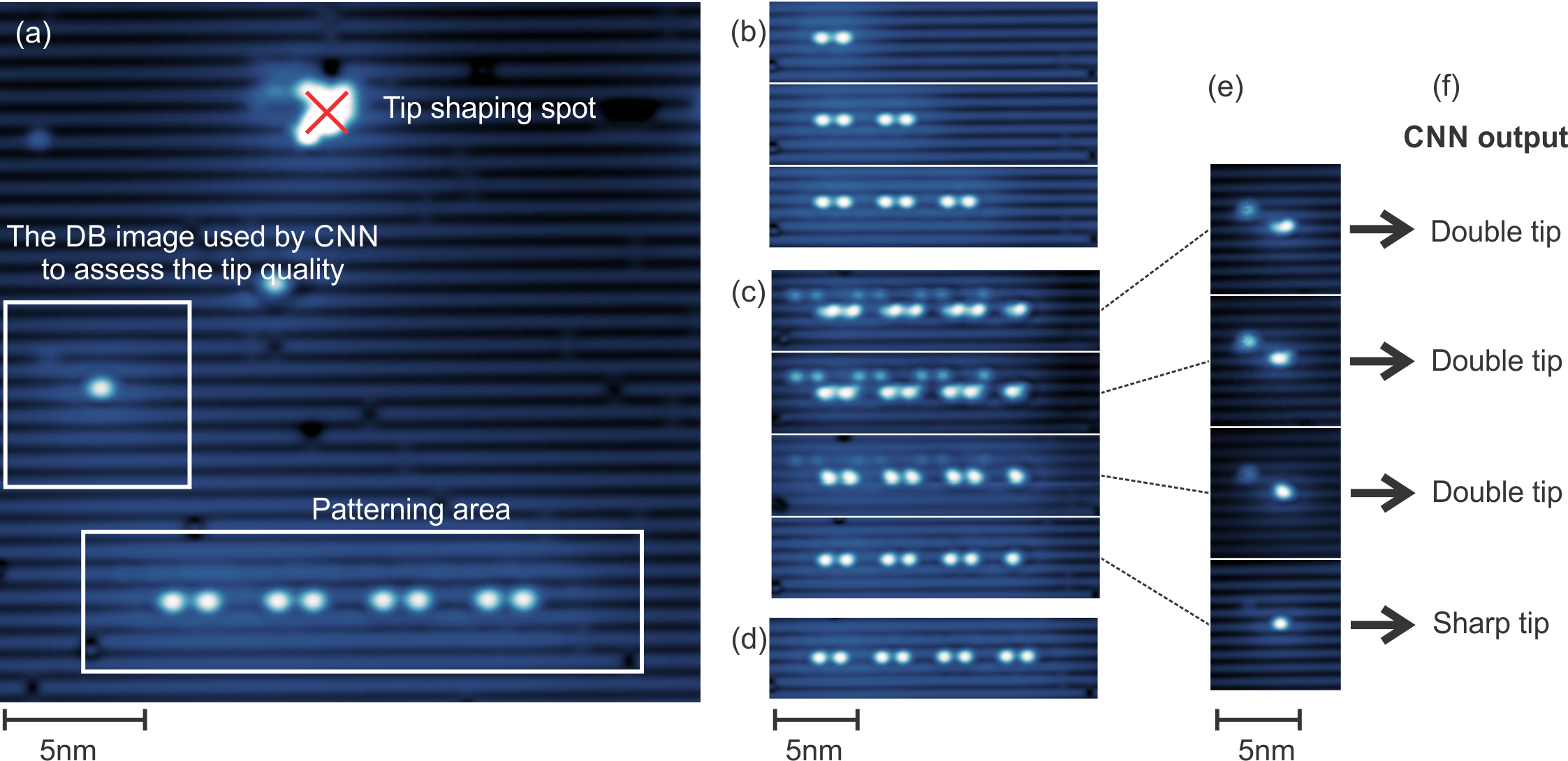}
	\caption{\label{Fig3} \textbf{Autonomous tip sharpening used along with atom-scale patterning.}  (a) An overview ($25\times 25$~nm$^2$) STM image showing a patterned binary atomic wire, an isolated DB used for tip quality assessment and a spot (red cross) chosen by the user to perform tip conditioning. (b) Sequence of patterning steps without noticeable tip quality change in between. (c) The tip became double after the creation of the last atom on the right and the user employed the tip conditioning routine to resharpen it. After three steps of automatic tip conditioning, the tip became sharp and the user carried on the pattering (d). (e) STM image of the isolated DB at middle left of (a) after each tip conditioning. The CNN used these images to assess the quality of the tip. (f) Output of the CNN for the images in (e). The tunneling conditions were $-1.8$~V and 50~pA for all images.}		
\end{figure*}
Figure~3a shows an STM image obtained with a ``double tip". Our program successfully identified the tip was not ideal and performed tip conditioning in an attempt to restore the tip's quality. Four conditioning steps were performed, at which point the program successfully recognized that the tip's quality had been restored (Fig.~3b) and the sequence was terminated. We note that the image frame to assess the quality of the tip must be carefully selected by the user to achieve accurate results. For instance, the defect close to the lower left dangling bond in Fig.~3b results in the outcome of ``double tip" for that dangling bond. Because all the other dangling bonds in the image frame are isolated, the program detect an accurate outcome.

As an example of how these techniques can be interfaced directly with existing automous atomic fabrication routines we demonstrate the use of our tip conditioning routine during the fabrication of a binary atomic wire from silicon dangling bonds~\cite{Huff2017a}. Figure~4a displays an STM image of the area that the experiment is performed. The user chose a defect-free area to fabricate the binary wire, an isolated DB to assess the tip quality using the  CNN, and a spot to perform tip indentation to recondition the tip. The wire was fabricated one atom at a time (Fig.~4b) by desorbing hydrogen atoms using voltage pulses (pulsed to $2.10$~V at the set-point of +1.3~V and 50~pA). We note that because the tip must be placed precisely over the hydrogren to be desorbed, double tips are unsuitable for atomic fabrication because they result in high patterning error rates. After making the seventh atom the quality of the tip unexpectedly deteriorated and the user initiated the tip-condtioning routine to fix it. The routine successfully detected the double tip by assessing the STM image of the isolated DB at middle left of Fig.~4a (assessments shown in Fig.~4e,f) and performed tip conditioning at the preselected spot. After three tip conditioning events the routine correctly identified that the tip quality had been restored and the user resumed fabrication. Figure~4d displays the final wire with eight atoms, where we note the superb image quality.

As a starting point for developing parallel atomic-precision fabrication tools we have implemented an automated routine that can evaluate the quality of an SPM probe and perform \textit{in-situ} conditioning to restore the quality of degraded tips. This routine is based on a CNN that assesses the probe quality by evaluating images of known atomic defects. As an example we trained the CNN with images of silicon dangling bonds, and demonstrated its use during the patterning of a binary atomic wire. The framework that we have developed here is an important step towards creating autonomous atom-scale fabrication tools and will be added as a module to the fully automated SPM patterning program we are currently developing~\cite{Achal2018}. This framework can easily be applied to other material systems and nanoscale imaging techniques. Applications other than atom scale fabrication, such as critical dimension analysis as used in modern semiconductor fabrication, will also benefit by variants of the techniques described here.

\begin{acknowledgments}
We would like to thank Mark Salomons and Martin Cloutier for their technical expertise.
We also thank  Alberta Innovates for their support.
M.R. thanks Wyatt Vine and Ken Gordon for their help to edit and proofread the manuscript. 
\end{acknowledgments}

\end{document}